\documentclass{ws-ijmpd}
\usepackage[super,compress]{cite}
\usepackage{subfigure}

\usepackage{hyperref}

\begin{document}
\markboth{Dehghani and Rahvar}
{Cross-Matching of OGLE III and GAIA catalogues: Investigation of Dark-lens microlensing candidates}

\title{Cross-Matching of OGLE III and GAIA catalogues: Investigation of Dark-lens microlensing candidates}

\author{Amirhosein Dehghani}
\address{Department~of~Physics,~Sharif~University~of~Technology,~P. O. Box~11365-9161,~Tehran,~Iran\\}
\author{Sohrab Rahvar}
\address{Department~of~Physics,~Sharif~University~of~Technology,~P. O. Box~11365-9161,~Tehran,~Iran\\
rahvar@sharif.edu}

\maketitle

\begin{history}
\end{history}

\newcommand{\RNum}[1]{\uppercase\expandafter{\romannumeral #1\relax}}


\begin{abstract}
In this work, we use $13$ microlensing candidates with dark lenses from OGLE III  catalogue \cite{2016MNRAS.458.3012W} and cross-match them with the GAIA catalogue. We identify the microlensing source stars in the GAIA catalogue by comparing the coordinate and the magnitude of stars and using the proper motion and the parallax parameters of the source stars. Combining with the microlensing light curves as well as the microlensing parallax effect, we determine the mass and the distance of lenses from Earth. We conclude that the lens for this set of microlensing events for the massive lenses are blackhole and for the small lenses are neutron/white dwarf candidates.
\end{abstract} 




\section{Introduction}
Finding dark objects such as black holes, neutron stars and white dwarfs in the Milky Way galaxy has always been an important field of study in astronomy.  The gravitational microlensing is a unique tool for the detection of Massive Astrophysical Compact Halo Objects (MACHOs) as the dark matter  \cite{pac}. Although the possibility of MACHOs as a candidate for dark matter has been ruled out \cite{Lasserre}, the gravitational microlensing has been used later as an astrophysical tool for studying remote stars \cite{2015IJMPD..2430020R} as well as discovering the exoplanets around the lens stars \cite{2010arXiv1002.0332G}. Recently it is also been used for the discovery of compact objects in the cosmological scales \cite{kalantar}.

For the single-lens microlensing events, the Einstein crossing time as the physical and the characteristic time of the events can be measured from the light curve. On the other hand, the Einstein crossing time is a function of the lens mass, the relative lens-source velocities and the distance of the lens and source stars from the observer. In other words, there is a degeneracy between the lens parameters. The perturbation effects such as parallax effect \cite{par1,2003A&A...412...81R} and finite-size effect \cite{fs,2019arXiv190610589G} that can break partially the degeneracy between the lens parameters. The other possibility to break degeneracy is the astrometry during the microlensing events\cite{2005A&A...438..153R}. Telescopes such as GAIA with a high angular resolution can measure the displacement of the centroid of light during the microlensing events \cite{2019arXiv191102584K}. 

The astrophysical importance of measuring the mass of lenses from the microlensing observations is that (i) we can obtain the mass function of stellar objects for single and binary lens, unlike to the traditional method where the mass function only is obtained in the binary systems, (ii) we can determine the mass function of dark and faint objects such as brown dwarfs, black holes, neutron stars, and white dwarfs.  

In this work, we use the third generation of Optical Gravitational Lensing Experiment, OGLE III \footnote{http://ogle.astrouw.edu.pl/} \cite{2008AcA....58...69U} for the photometric light curves, operated from 2001 until 2009, located at the Las Campanas Observatory. In addition, we use the astrometric data of GAIA to extract additional physical parameters of the microlensing systems. GAIA is a European Space Agency (ESA) mission \cite{2016A&A...595A...1G} designed for full-sky astrometry of stars in the Galaxy. So far, they have released two series of data in which they reported celestial positions and the apparent magnitude for approximately 1.7 billion sources \cite{2018A&A...616A...1G}. For 1.3 billion of those sources, parallaxes and proper motions have been measured. Here, we cross-match GAIA and OGLE catalogues and use the proper motion and the distance of the source stars from GAIA database to have extra parameters of the microlensing events. We use a sub-sample of 13 microlensing events where their lenses are dark and the source star is least blended. For this set of events, we assume them as dark lenses. Combining the microlensing-parallax effect from the light curve and the proper motion and parallax measurements from the GAIA, we obtain the mass of lenses for this set of events with higher accuracy. Using the GAIA catalogue in the microlensing observations opens a new window for breaking the degeneracies in the lens parameters. 

      
The outline of this paper as is follows: In section (\ref{GM}) we introduce the theory of gravitational microlensing and degeneracy problem. In section (\ref{GAIA}), we use the GAIA data from the space-based telescope and cross-match it with OGLE III catalogues. Also, we investigate the feasibility of using extra data for the microlensing events from the GAIA catalogue. In section (\ref{result}) we choose a sub-sample of the microlensing events where the lenses are non-luminous. From the astrometry and proper motion of the source star in the GAIA data, we calculate the mass and distance of lenses with higher accuracy.  The conclusion is given in section (\ref{conclusion}). 

\section{Gravitational Microlensing and Degeneracy problem}
\label{GM}
Light bending, as the prediction of the general relativity, is the deflection of the light rays around a massive object. From the 
Schwarzschild metric, the deflection angle is given by $\alpha= 4GM/bc^2$ \cite{1936Sci....84..506E}, where $M$ is the mass of the lens, and $b$ is the impact parameter of the light ray as the closest distance of the light from the lens. For the gravitational microlensing of stars inside the Milky Way galaxy, the angular separation between the images are smaller than the angular resolution of the ground-based telescopes. The effect of lensing is the magnification of the background star during the lensing. 
 
The characteristic angular size of the lens is given by the Einstein angle \cite{1992grle.book.....S} as follows:
\begin{equation}
\label{eq:eangle}
\theta_E=0.901mas(\frac{M}{M_{\odot}})^{1/2}(\frac{D_S}{10kpc})^{-1/2}(\frac{1-x}{x})^{1/2},
\end{equation}
where $M$ is the lens mass, $D_S$ is the source distance, and $x$ is the lens distance divided by the source distance (i.e. $x = D_L/D_S$). Multiplying this parameter to the lens distance is the Einstein radius,  $R_E$.

Since inside the Galaxy, the relative position of the Earth and other stars due to the internal dynamics of the Galaxy change over time, we would expect that the lensing configuration also changes over time. For a single lens, the relative velocity of the lens with respect to the source star is a straight line. The associated time-scale of lensing is the crossing of the source star across the Einstein angle (i.e. $t_E = \theta_E/\mu$), so-called the Einstein crossing time, and is given by
 \begin{eqnarray}
\label{eq:etime}
t_E=45.6day(\frac{D_S}{8.5kpc})^{1/2}(\frac{M}{0.5M_{\odot}})^{1/2}(\frac{1-x}{x})^{1/2}\cr\times(|V_{S,\bot}-V_{E,\bot}-\frac{1}{x}(V_{L,\bot}-V_{E,\bot})|\frac{1}{220km/s})^{-1},
\end{eqnarray}
where $V_{S,\bot}$, $V_{L,\bot}$, and $V_{E,\bot}$ are the transverse velocities of source, lens, and Earth with respect to our line of sight to the source star, respectively.  The major problem with the standard microlensing events is that we have only one physical parameter of $t_E$ from our measurements and we can not extract lens parameters as the mass, the distance of lens and source, and the relative velocities of the lens and source from this single parameter.

The gravitational microlensing has been used for detecting compact massive objects inside the Galactic halo as well as the discovery of the exoplanets orbiting around the lens stars \cite{2010arXiv1002.0332G}.
 In these observations, survey telescopes such as OGLE, MOA, and KMTNet monitor millions of stars in the Galactic bulge. 
 In addition to these telescopes, the follow-up telescopes monitor the microlensing events with high cadence to identify exoplanet signatures in the light curves. 
 Also the individual events are observed by Hubble, Spitzer, and Keck telescopes as the follow-up observation \cite{2001ApJ...552..582A,2019AJ....157..106S,2015ApJ...808..170B}  and will be observed by future space-based telescopes as 
WFIRST  \cite{2019ApJS..241....3P,Bagheri_2019}.

\subsection{Parallax and Xallarap}
The orbital motion of the earth around the sun results in a perturbation in the microlensing light curve, so-called microlensing parallax effect \cite{par1,2003A&A...412...81R}. In the parallax effect, the relative velocity of the source with respect to the lens deviates from the straight line. From the parallax parameter in the microlensing, we can obtain extra information about the lens.  For a microlensing event the parallax parameter is given by 
\begin{equation}
\label{eq:mparallax}
\pi_E=\frac{a_\oplus(1-x)}{R_E},
\end{equation}where $a_\oplus=1A.U.$ is the distance of earth to sun. Here, $\pi_E = \pi_E(x,M,D_S)$ is a function of lens mass and distance of lens and distance of source star from the observer. If we fix the distance of the source star, then we can constrain the mass and distance of the lens from the observer.  There is another perturbation effect, so-called Xallarap effect \cite{2008mmc..confE..37R} which is similar to parallax, but instead of considering observer motion around the sun, the orbital motion of source star around its companion produces an asymmetric effect in the light curve.
\subsection{Finite-source effect}
One of the assumptions of the standard microlensing event is that the source star is point-like. For a relative close impact parameter of the lensing compare to the angular size of the source star, the finite-size effect is important.  In the finite-size effect different parts of the source star have different impact parameters and get different magnifications. The result is the change in the shape of the light curve at the peak. Using the parallax and finite-source effects in light curves of microlensing curves, we can break partially the degeneracy between the parameters of the lens \cite{2012ApJ...751...41C,2019arXiv190610589G}.

\section{GAIA and OGLE III catalogues cross-match}
\label{GAIA}
For microlensing data, we use OGLE \RNum{3} data which is produced from the year 2002 to 2009 \cite{2008AcA....58...69U}. We choose OGLE III data because of the photometric accuracy in the light curves, also the microlensing events for this set of data have happened a long time before observations of GAIA where  during the time, the lens and the source stars might be resolved by the GAIA telescope.  
 
For this set of microlensing events, we can also search for lens and source stars with the ground-based adaptive optics  such as the Keck telescope that could resolve the lens and source star for the event of OGLE-2005-BLG-071L \cite{Bennett_2020}. Recently follow-up observation of a long duration microlensing event by Hubble Space Telescope revealed that lens star is a blackhole \cite{2022arXiv220113296S}. Here,  we use GAIA data and investigate the microlensing events by cross-matching it with the OGLE catalogue to extract possible information for the proper motion and distance of the lens and source stars. In the case of identifying the source and the lens stars in GAIA catalogue, we can use the proper motion and the distance of lens and source in the right-hand side of equation (\ref{eq:etime}) and the Einstein crossing time at the left-hand side of this equation from the microlensing light curve. The result would be identifying the mass of the lens from this equation.  
 
Taking into account the angular resolution of GAIA, it needs enough time after microlensing to resolve both lens and source stars. Here in this work, we analyze a sub-set of events with possible dark lenses. These events have the least blending and on the other hand, have long duration events \cite{2016MNRAS.458.3012W} with the microlensing parallax signal. Details on our candidates will be discussed in the next section.

\begin{figure}	
	\includegraphics[width=\columnwidth]{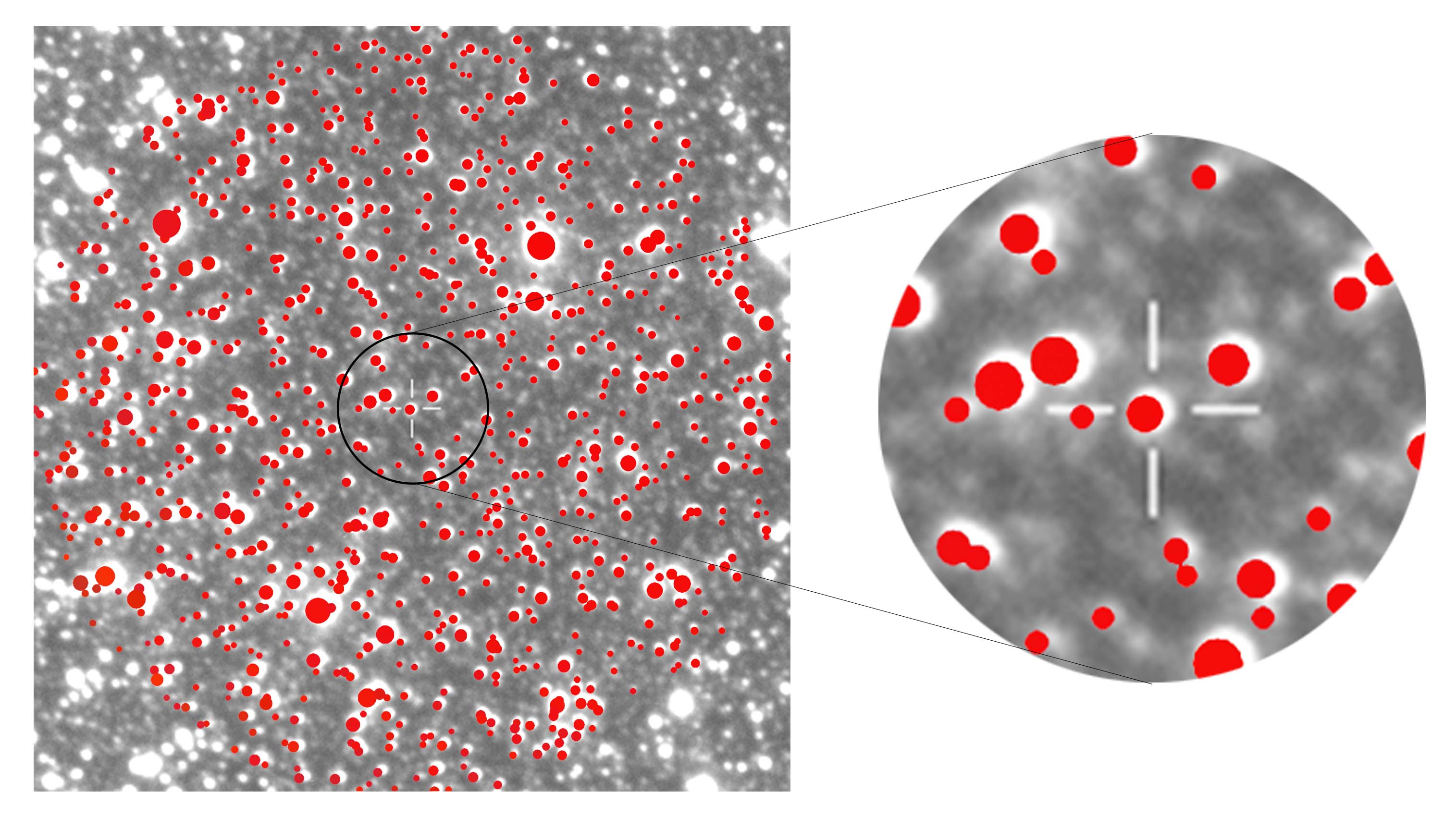}
	\caption{OGLE 2006-BLG-095 event crossmatched with GAIA, representing a sample of our cross-match between OGLE III and GAIA catalogues. North is up and East is to the left. The microlensing source star is assigned by cross sign.}
	\label{fig:example}
\end{figure}

We note that GAIA FITS images will not be available until the end of the mission in 2022 and in order to identify the GAIA stars, according to the coordinate and the magnitude of the stars, we can generate artificial images from a given field. 
In the second step, GAIA sources are superimposed on the OGLE III images. Figure (\ref{fig:example}) shows a sample of our cross-match between OGLE III and GAIA catalogues for OGLE 2006-BLG-095 event. The red spots are the GAIA sources that are synthetically generated according to the location of each star where the size of the spot is proportional to the brightness of the star.

\begin{figure}	
	\includegraphics[width=\columnwidth]{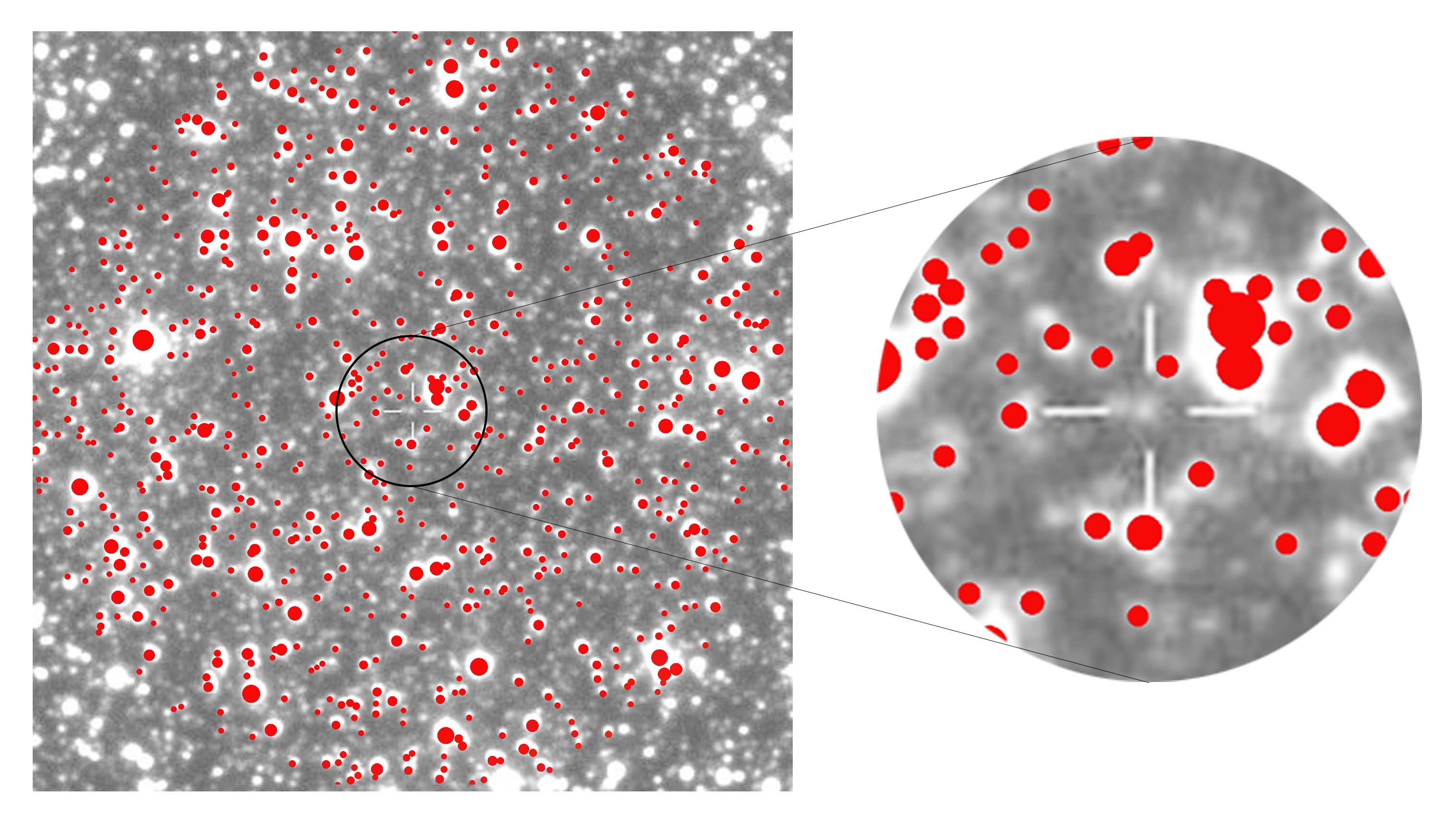}
	\caption{Crossmatch between the OGLE III and GAIA catalogues. For OGLE 2005-BLG-036, the corresponding star in the GAIA catalogue is not identified. North is up and East is to the left.}
	\label{fig:oddexample}
\end{figure}

The source star for the microlensing is identified by a cross sign. We can also make this crossmatching with the Aladin sky Atals \cite{aladin,aladin2}. For some of the events such as OGLE 2005-BLG-036  in Figure (\ref{fig:oddexample}), while all the stars around the microlensing source star are identified in GAIA catalogue, the source star is missing in the  catalogue. Also, we have checked the source star of this microlensing event with the 2MASS catalogue where this star is also not identified. So missing some of the source stars might be due to high magnification of faint stars during the microlensing where it is not identified in the other catalogues.

\subsection{How GAIA improves the results}
In order to solve the degeneracy in equation (\ref{eq:etime}), the conventional method is 
using the distribution functions for the velocities of the lens and source stars based on well-known galactic models and compute an overall distribution for the relative angular velocity which depends on distances of the lens and the source stars as follows 
\begin{equation}
\label{mu}
\mu = 5.46 \text{mas/yr}(\frac{8.5 \text{kpc}}{D_S})\frac{|V_{S,\bot}-V_{E,\bot}-\frac{1}{x}(V_{L,\bot}-V_{E,\bot})|}{220 km/ s}. 
\end{equation}
The astrometric observations of lens and source stars by GAIA telescope enable us to measure directly the proper motion and distance of stars. However, we note that the GAIA resolution is well below the characteristic relative separation of lens and source stars (see table 3 in \cite{2018A&A...616A...1G}). In order to resolve the lens and source stars, astrometric observations has to be done few years after the peak of the microlensing events. Taking into account the angular resolution of the stars in GAIA which is in the order $\sim100$ mas, we can not resolve the lens and the source stars after few year, except for high velocity lens or source stars. We will discuss details of this issue in section (\ref{sec:possibility}). Here, we will investigate a small sub-class of the dark lenses where in the GAIA catalogue we can follow  the proper motion of only source star. 

Since most of microlensing events have been observed in the galactic bulge direction, a reasonable assumption is that the source stars of microlensing events are in galactic bulge. This is a reasonable assumption for studying a large number of microlensing events, but for  the individual events that is highly uncertain. In addition, assuming the source stars being in galactic bulge with an isotropic dispersion velocity of the stars in the bulge following a Gaussian velocity provides a large uncertainty in the source velocity.  For a sub-class of microlensing events with the dark lens, we use GAIA catalogue to identify the distance and the proper motion of the source stars. We will analyze these events in section (\ref{result}). 
 
\subsection{Possibility of detecting both lens and source stars of microlensing events in GAIA catalogue}
\label{sec:possibility}

We investigate the possibility of detecting both the lens and source stars of microlensing events in the GAIA catalogue. During the time that the microlensing event is happening, the lens and the source stars are almost along the same line of sight. However, as a result of relative proper motion, we would expect that the lens and source stars can be resolved after a while. In order to estimate the feasibility of resolving both the lens and source stars in GAIA catalog, we take a sub-sample of GAIA data towards the galactic bulge with the direction of Ra$=17:45:40.04$ and Dec $= -29:00:28.1$, and the angular radius of $7$ degrees. 

We choose stars with the parallax parameter to be in the range of $\pi \in [0.10, 0.18]$ mas, equivalent to the distance of stars between $5.5$ kpc and $10.0$ kpc with respect to the earth. This distance covers  the galactic bulge along our line of sight. We exclude  $68.4\%$ of stars in our list due of large error in parallax measurement ($\sigma_\pi/\pi>1$). The result is $1149367$ stars in this direction with reasonable accuracy in the distance measurement. From the observed proper motion of stars, we obtain a distribution function for the proper motion of stars in the bulge. This is a Gaussian function with the mean velocities of  $\mu_{RA} = -2.5$ mas/yr and  $\mu_{Dec} = -4.8$ mas/yr in RA and Dec directions and corresponding variance of $\sigma_{RA}\simeq  2.6$ mas/yr and $\sigma_{Dec} = 2.8$ mas/yr,  respectively .
	
We assume that the lens stars are too faint to be resolved in the Galactic bulge and from the theory of microlensing most  the lenses should be located in the Galactic disk \cite{2017A&A...604A.124M}. So the distance and proper motion of the sub-sampled of stars we choose in the GAIA catalogue can be considered as the source stars for the microlensing events.  For generating a proper motion of the lens stars, we use the probability distribution of lens being located at the distance of $x$ by $dp/dx\propto\rho(x)\sqrt{x(1-x)}$. Here we assume that each lens while moving on the lens plane produces a tube with the width of $2R_E$ and the length of $v_t T$ where $v_t$ is the transverse velocity and $T$ is the duration of observation. So the Probability of detection of microlensing events which is the area covered by the lens is proportional to the Einstein radius \cite{2015IJMPD..2430020R}. We also use the distribution function for the velocity of the lenses which is a combination of the global velocity and the dispersion velocity of the disk \cite{2017A&A...604A.124M,2009A&A...500.1027R,2011AN....332..461B}. The global rotation of the disk is given as a function of the galactocentric distance by
\begin{equation}
\label{vrot}
V_{rot}(r)=V_{rot,\odot}[1.00767(\frac{r}{R_\odot})^{0.0394}+0.00712],
\end{equation}
where r is in cylindrical coordinates and $V_{rot,\odot} = 239\pm 7$~km/s and $R_\odot =8.5$ kpc. Moreover, the peculiar velocity of the stars of disk is described
by an anisotropic Gaussian distribution with the following
radial, tangential, and perpendicular velocity dispersions \cite{2012A&A...547A..71P}:
\begin{eqnarray}
\label{sigma}
\sigma_r=27.4\pm1.1~ \text{km/s}, \cr
\sigma_\theta=20.8\pm1.2~ \text{km/s}, \cr
\sigma_z=16.3\pm2.2~ \text{km/s}.
\end{eqnarray}

\begin{figure}
	\includegraphics[width=\columnwidth]{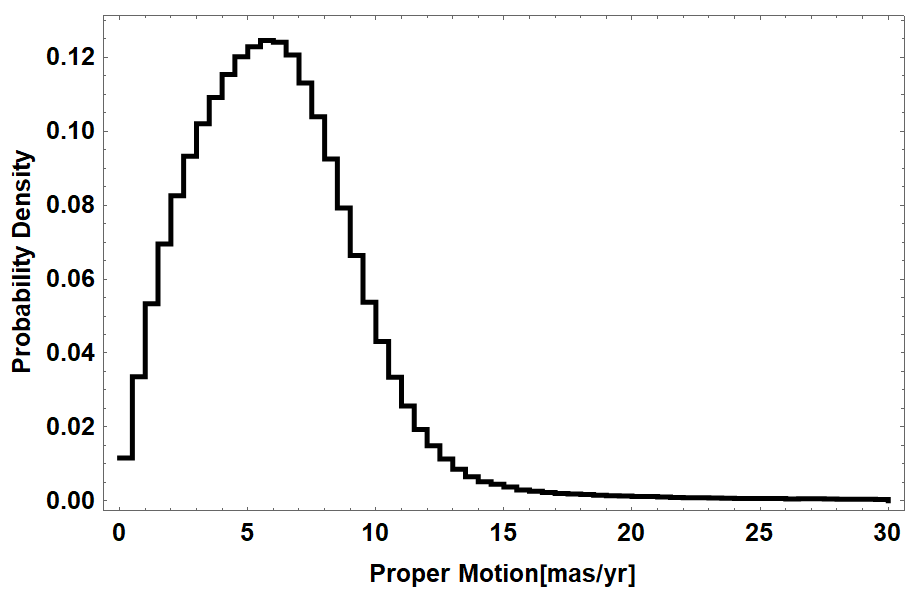}
	\caption{Distribution of relative proper motion of the source with respect to the lens stars in the microlensing events toward the Galactic bulge, resulting form the simulation and using GAIA data.}
	\label{fig:drpm}
\end{figure}

\begin{figure}
	\includegraphics[width=\columnwidth]{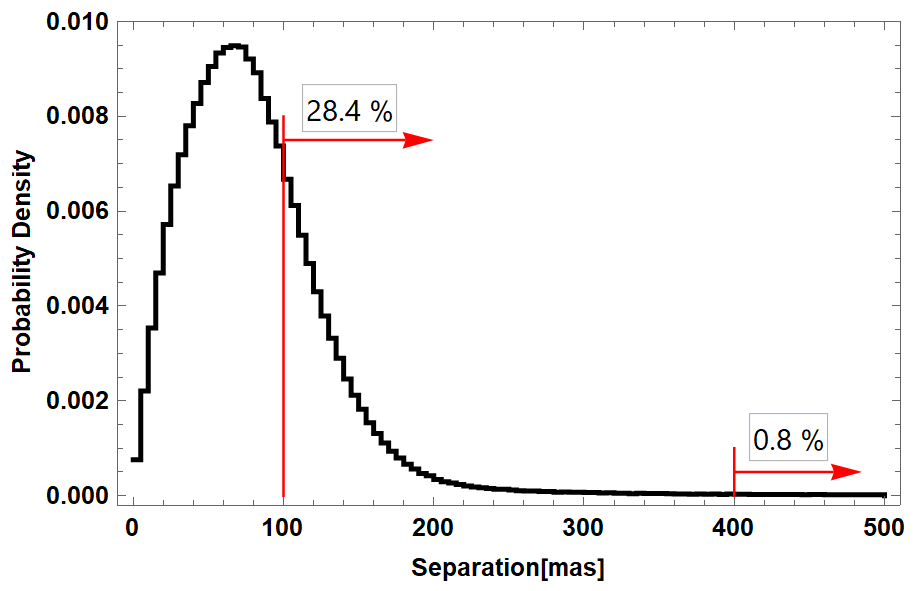}
	\caption{Expected distribution for the angular separation of the lens and source stars of the microlensing events from OGLE III, observed in the GAIA catalogue. For $28.4\%$ of stars in GAIA catalogue where the distance of stars is well measured, the angular separation is larger than 100 mas and for $0.8\%$ of stars the angular separation is larger than $400$ mas.}
	\label{fig:drs}
\end{figure}

Combining the location of lens and the relative velocity of the lens with respect to the Earth, we calculate the proper motion of the lens with respect to the observer. 
 
In our Monte-Carlo simulation, we combine the relative proper motion of the lens (from the simulation) and the relative proper motion of the source stars (from the GAIA data).  Figure (\ref{fig:drpm}) represents the distribution function for the relative proper motion of the lens with respect to source stars in terms of mas/yr. In order to calculate the separation of lens and source stars in the GAIA catalogue, we take the observational duration of OGLE  from the year 2001 to 2009. Since the observation of GAIA is done in 2018,  there are 9-17 years time-separation between these two surveys.
In order to compare the separation between the lens and source stars with the observational resolution of GAIA, we generate synthetic angular distance between lens and source stars. We fix the rate of annual microlensing events from years 2001-2009, in our simulation and find the time-separation of $\Delta T = t_{GAIA}-t_{OGLE}$ (representing the time between the two observations). Multiplying this time to the relative proper motion of lens and source  from Figure (\ref{fig:drpm}) results in the angular separation between the lens and source stars.

\begin{table*}
\tbl{Parameters used for calculating masses and distances of all 13 events with dark lens. The first block is data from microlensing observations and the second block from the GAIA observations.}
	{
	\begin{tabular}{|l|l|l|l|l|l|l|l|l|l|}
		\hline
     	\multicolumn{5}{|l|}{\textbf{Microlensing Data}} & \multicolumn{4}{l|}{\textbf{GAIA Data}} \\ \hline
		EWS-OGLE & $u_0$ & $t_E^{helio}$ & $\pi_{EN}$ & $\pi_{EE}$ & GAIA source ID & Parallax & $\mu_{Ra}$ & $\mu_{Dec}$ \\
		(or OGLE DB ID) & sign & [days] &  &  &  & [mas] & [mas/year] & [mas/year] \\ \hline\hline
		  {*}{2006-BLG-095} & $+$ & $296.1_{-7.4}^{+7.6}$ & $+0.033_{-0.001}^{+0.001}$ & $-0.051_{-0.002}^{+0.002}$ &   {4062558472490201728} &   {$0.052\pm0.167$} &   {$-4.53\pm0.26$} &   {$-6.45\pm0.20$} \\ 
		& $-$ & $255.9_{-5.3}^{+7.4}$ & $-0.033_{-0.001}^{+0.001}$ & $-0.073_{-0.003}^{+0.004}$ &  &  &  &  \\ \hline
		  {*}{BLG342.5.73806} & $+$ & $184.0_{-3.2}^{+4.0}$ & $+0.125_{-0.003}^{+0.003}$ & $+0.165_{-0.005}^{+0.005}$ &   {4068855482936627200} &   {$0.260\pm0.079$} &   {$-0.40\pm0.13$} &   {$-1.80\pm0.10$} \\ 
		& $-$ & $185.2_{-3.2}^{+4.0}$ & $-0.128_{-0.003}^{+0.003}$ & $+0.154_{-0.005}^{+0.005}$ &  &  &  &  \\ \hline
		  {*}{2004-BLG-361} & $+$ & $174.1_{-3.7}^{+5.0}$ & $+0.074_{-0.002}^{+0.002}$ & $-0.041_{-0.002}^{+0.002}$ &   {4041821408382645376} &   {$0.198\pm0.091$} &   {$-0.21\pm0.13$} &   {$-3.94\pm0.11$} \\ 
		& $-$ & $175.5_{-4.5}^{+4.7}$ & $-0.084_{-0.003}^{+0.003}$ & $-0.060_{-0.003}^{+0.003}$ &  &  &  &  \\ \hline
		  {*}{2005-BLG-474} & $+$ & $137.5_{-1.8}^{+1.8}$ & $-0.074_{-0.009}^{+0.010}$ & $-0.089_{-0.006}^{+0.006}$ &   {4056324241497959296} &   {$0.232\pm0.141$} &   {$-0.46\pm0.20$} &   {$-1.72\pm0.17$} \\ 
		& $-$ & $147.9_{-1.4}^{+1.4}$ & $+0.042_{-0.014}^{+0.012}$ & $-0.070_{-0.005}^{+0.005}$ &  &  &  &  \\ \hline
		2005-BLG-351 & $+$ &  $135.4_{-5.8}^{+9.5}$ & $-0.236_{-0.010}^{+0.010}$  & $-0.037_{-0.014}^{+0.016}$ & 4041737712212920192& $-0.666\pm0.401$ & $-8.65\pm0.55$ &$-7.01\pm0.45$  \\ \hline
		BLG172.8.614& $+$ &  $159.1_{-4.4}^{+5.3}$ & $-0.199_{-0.005}^{+0.005}$  & $-0.151_{-0.005}^{+0.005}$ & 4043666182743831936& $-0.014\pm0.329$ & $-2.69\pm0.43$ &$-5.15\pm0.37$  \\ \hline
		  {*}{2002-BLG-061} & $+$ & $171.7_{-8.8}^{+11.8}$ & $-0.035_{-0.047}^{+0.052}$ & $+0.058_{-0.005}^{+0.004}$ &   {4061317643505236352} &   {$0.931\pm0.666$} &   {$-6.04\pm1.06$} &   {$-3.39\pm0.78$} \\ 
		& $-$ & $170.6_{-6.1}^{+7.1}$ & $+0.060_{-0.060}^{+0.050}$ & $+0.062_{-0.004}^{+0.005}$ &  &  &  &  \\ \hline
		2008-BLG-096 & $+$ &  $138.2_{-5.3}^{+5.6}$ & $-0.036_{-0.017}^{+0.016}$  & $-0.140_{-0.006}^{+0.006}$ & 4063554084450730880& $0.276\pm0.235$ & $+2.53\pm0.62$ &$-0.65\pm0.48$  \\ \hline
		2005-BLG-372 & $+$ &  $91.4_{-1.7}^{+2.4}$ & $-0.069_{-0.015}^{+0.012}$  & $+0.036_{-0.009}^{+0.007}$ & 4056570325872878720& $-1.053\pm0.296$ & $-1.48\pm0.40$ &$-6.47\pm0.35$  \\ \hline
		2005-BLG-086 & $+$ &  $95.2_{-2.2}^{+2.3}$& $+0.245_{-0.034}^{+0.031}$  & $+0.110_{-0.008}^{+0.008}$ & 4063251615722434176& $-0.264\pm0.148$ & $-2.23\pm0.27$ &$-3.27\pm0.22$  \\ \hline
		  {*}{2005-BLG-020} & $+$ & $69.8_{-1.7}^{+2.4}$ & $+0.152_{-0.036}^{+0.052}$ & $+0.126_{-0.006}^{+0.006}$ &   {4050127570081324672} &   {$-0.087\pm0.090$} &   {$-2.63\pm0.16$} &   {$-6.10\pm0.15$} \\ 
		& $-$ & $67.4_{-0.8}^{+0.8}$ & $-0.256_{-0.041}^{+0.040}$ & $+0.113_{-0.006}^{+0.006}$ &  &  &  &  \\ \hline
		2006-BLG-251 & $+$ &  $80.5_{-1.1}^{+1.1}$ & $+0.210_{-0.047}^{+0.041}$  & $-0.065_{-0.009}^{+0.009}$ & 4062443229854120320& $-0.014\pm0.170$ & $-5.33\pm0.29$ &$-6.08\pm0.22$  \\ \hline
		  {*}{2008-BLG-013} & $+$ & $78.5_{-0.9}^{+1.0}$ & $-0.066_{-0.074}^{+0.068}$ & $-0.038_{-0.009}^{+0.008}$ &   {4056077199264203648} &   {$0.083\pm0.164$} &   {$+0.18\pm0.28$} &   {$-0.58\pm0.22$} \\ 
		& $-$ & $78.3_{-0.8}^{+0.9}$ & $-0.071_{-0.083}^{+0.077}$ & $-0.038_{-0.009}^{+0.008}$ &  &  &  &  \\ \hline
	\end{tabular}
	}
		\label{tab:inserts}
\end{table*}

Figure (\ref{fig:drs}) shows the result of our simulation. Noting that the angular resolution of GAIA is about $100$ mas  \cite{2018A&A...616A...2L},  we would expect that in $28.4\%$ of microlensing events we can resolve the lens and source stars from a subset of GAIA catalogue with reasonable parallax error. If we take into account all the stars of GAIA catalogue in the direction of Galactic bulge, for $32\%$ of stars with reasonable measured parallax parameter, the percentage of source-lens stars can be resolved is $28.4\%\times 0.32 =  9.10\%$. In practice, investigating the GAIA data, the minimum separation of stars is not less than $400$ mas. If we set this number as the angular resolution of stars in GAIA, from Figure (\ref{fig:drs}),  $0.8\%$ of events or in terms of all stars in our study or in terms of whole GAIA catalogue  $0.80\%\times 0.32 = 0.26\%$ of microlensing events with lens and source stars can be resolved.  The Keck telescope data could resolve the lens and source stars of  OGLE-2005-BLG-169 event with the angular separation of 61 mas \cite{2015ApJ...808..170B}. This angular separation is around the peak of Figure (\ref{fig:drs}) as the most probable angular separation from the simulation. In this work, we will not compare the whole events of OGLE with the GAIA catalogue and choose only microlensing events expecting the lens of events are dark. 

\section{Mass and distance analysis of 13 dark lens candidates }
\label{result}
We have seen that the chance of resolving the lens and the source stars in the GAIA catalogue ($>400$ mas) is low. 
Here we investigate  $59$ microlensing events which exhibit strong parallax effect and have been studied in \cite{2016MNRAS.458.3012W}. The position of these stars in the color magnitude diagram shows that they are more likely located at the bulge which we can fix source distance at $8$~kpc. Then, by considering the most probable proper motion between source and lens based on a galactic model, they calculate the mass and the distance of the lenses. Finally, by comparing blending parameter and the mass of lens inferred from the parallax microlensing effect, they associate 
an apparent magnitude to the lens stars. Out of $59$ events with the strong parallax signature, they identity $13$ of them might have dark lenses. Table (\ref{tab:inserts})
is the list of the 13 microlensing candidates with possible dark lenses  where from the MCMC positive and negative solutions for $u_0$, the Einstein crossing time and microlensing parallax vector $(\pi_{EN},\pi_{EE})$ are reported

\begin{figure*}
	\label{all}
	\includegraphics[width=150mm]{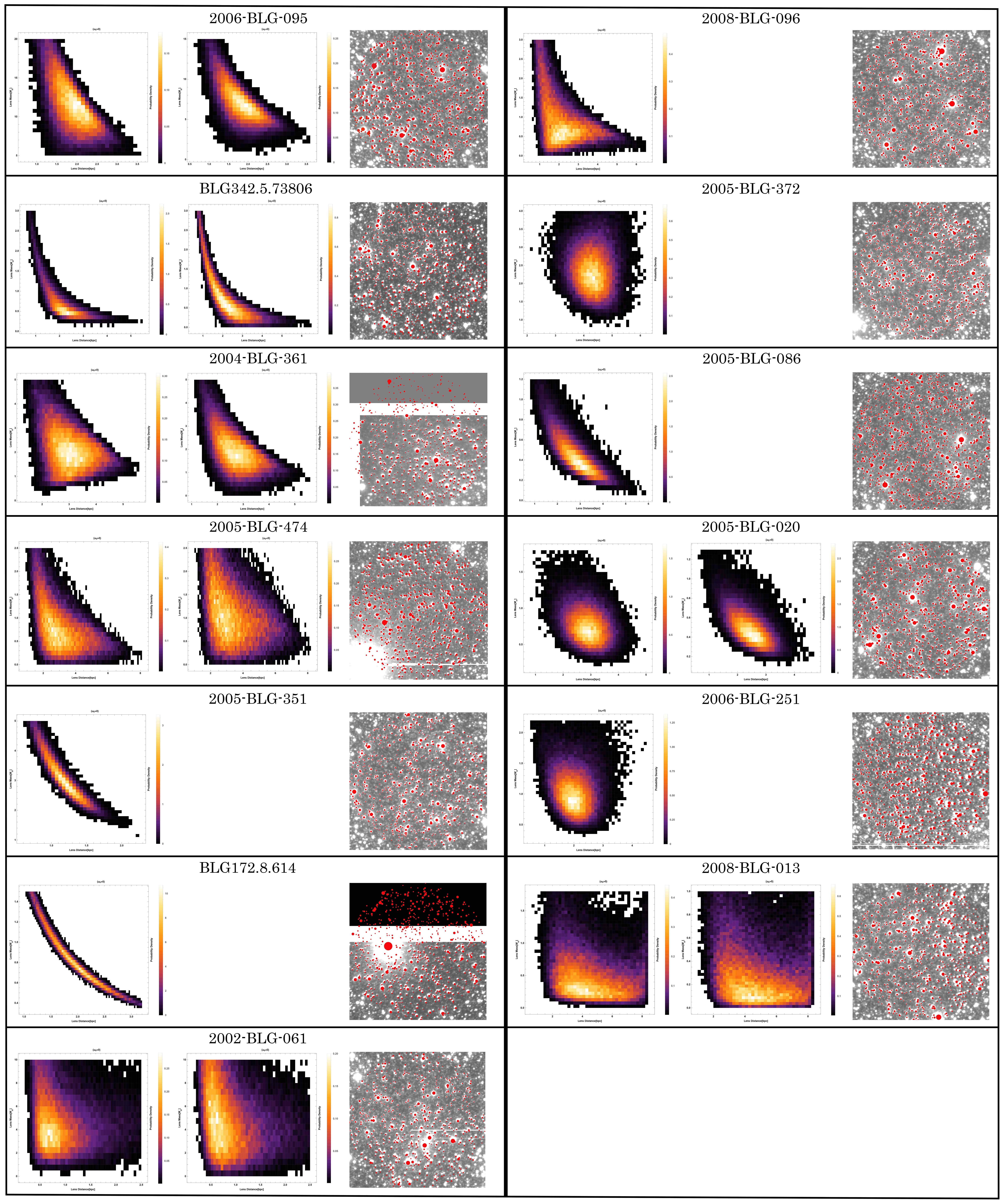}
	\begin{center}
	\caption{The result from analysis of 13 microlensing events. For each event (OGLE ID is written on top of chart) the cross-match between the OGLE and GAIA catalogue is represented (North is up and East is to the left in these figures) in the third panel for event from right side. The probability function for the mass (in $y$ axis) and distance of lens (in $x$ axis) is represented by two dimensional histogram. The first panel for each event is for the positive $u_0$ and the second panel is for negative $u_0$.}
	\end{center}
\end{figure*}

We cross list these 13 events with the GAIA catalogue. The overlap between the 
finding chart of the OGLE catalogue and GAIA is shown in Figure (5). For each of the microlensing source stars, we find the corresponding star in the GAIA catalogue. We also cross checked the apparent I-band magnitude of the source star in OGLE with G$_{RP}$ magnitude in GAIA.  Out of $13$ stars,  $7$ stars are reported with the positive parallax parameter in the GAIA catalogue  \cite{2018A&A...616A...9L} while for all the $13$ stars GAIA gives the proper motion of the source star. We note that the observed proper motion is the relative motion of the source with respect to the earth, i.e. $\mu = (V_{S,\bot} - V_{E,\bot})/D_S$. 

For $7$ events, we have the distance and the proper motion of the source stars. Since there is an uncertainly in the distance of source stars from the GAIA measurements, we impose extra constrain such that the source stars are not  located at the other side of the Bulge. In what follows, we use the extra data from the GAIA catalogue to confine the lens parameter with a better accuracy.  We use the two parameters of the Einstein crossing time ($t_E$) and the parallax parameter $(\pi_E)$ from analyzing the microlensing light curves. The Einstein crossing time from equation (\ref{eq:etime}) is a function of the source distance, lens distance, lens mass and the relative proper motion between lens and source stars. Also the microlensing parallax parameter from equation (\ref{eq:mparallax}) is a function of source distance, lens distance, and the lens mass. 

We use the total parallax parameter of $\pi_E=\sqrt{\pi_{EN}^2+\pi_{EE}^2}$ in equation (\ref{eq:mparallax}) in which $\pi_E = \pi_E(x,M,D_S)$.  For a fixed $D_S$ with the corresponding error bar we put constrain between the mass of the lens  (i.e. $M$) and distance of the lens from the observer(i.e. $x$ parameter). Substituting this constrain in the Einstein crossing time, equation (\ref{eq:etime}), we replace the mass of lens from the parallax constrain with the distance of the lens and find another constrain between the relative velocity of lens-observer-source and the distance of the lens from the observer.  
Now, we choose a distribution function for this relative velocity from the Galactic model and obtain a likelihood distribution function for the distance and the mass of the lenses, see Figure (5). Table (\ref{tab:results table}) compared our results with the recent work in  \cite{2016MNRAS.458.3012W}.
For $6$ events out of $13$ event, we don't have the distance of the source star where we fix the distance of source stars for these events at $D_S = 8.5$kpc. For these events we use the  proper motion of the source stars from GAIA catalogue.



\begin{table*}
	\tbl{The mass and distance for 13 dark-lens events deterimined with $1\sigma$ error. The distance of source stars of those microlensing events that is not identified is fixed to $8.5$ kpc. The results from our analysis is compared with \cite{2016MNRAS.458.3012W}.}
	{
	\begin{tabular}{|l|l|l|l|l|l|l|l|l|}
		\hline
		\multicolumn{2}{|l|}{} & \multicolumn{3}{l|}{\textbf{Our Results}} &\multicolumn{2}{l|}{\textbf{Previous Results}} \\ \hline
		EWS-OGLE & $u_0$ & source distance & lens mass& lens distance & lens mass& lens distance \\
		(or OGLE DB ID) & sign & [kpc] & [solar mass] & [kpc]& [solar mass]& [kpc] \\ \hline\hline
		  {2006-BLG-095} & $+$ &   {$5.1_{-1.8}^{+2.6}$ } & $11.4_{-2.9}^{+4.1}$ & $1.81_{-0.41}^{+0.44}$ & $9.3_{-4.3}^{+8.7}$& $2.4_{-1.0}^{+1.1}$	 \\ & $-$ & & $6.6_{-1.9}^{+2.9}$ & $1.75_{-0.42}^{+0.45}$ & $8.7_{-4.7}^{+8.1}$& $1.8_{-0.8}^{+1.1}$ \\ \hline
		  {BLG342.5.73806} & $+$ &    {$3.8_{-0.9}^{+1.5}$ } & $0.61_{-0.19}^{+0.68}$ & $2.03_{-0.69}^{+0.70}$ & $1.3_{-0.6}^{+1.5}$& $1.7_{-0.8}^{+1.2}$ \\ & $-$ & & $0.65_{-0.37}^{+1.08}$ & $2.1_{-0.9}^{+1.0}$ & $1.6_{-0.8}^{+1.7}$& $1.6_{-0.8}^{+1.1}$ \\ \hline
		  {2004-BLG-361} & $+$ &   {$4.6_{-1.3}^{+2.1}$ } & $2.05_{-0.70}^{+0.94}$ & $2.95_{-0.63}^{+0.75}$ & $3.3_{-1.5}^{+2.7}$& $2.9_{-0.9}^{+1.1}$\\ & $-$ & & $1.83_{-0.66}^{+0.93}$ & $2.65_{-0.62}^{+0.72}$ & $4.8_{-2.5}^{+4.0}$& $1.8_{-0.7}^{+1.2}$ \\ \hline
		  {2005-BLG-474} & $+$ &   {$3.8_{-1.2}^{+2.4}$ } & $0.73_{-0.37}^{+0.67}$ & $2.9_{-0.9}^{+1.4}$ & $3.1_{-1.6}^{+3.1}$& $2.1_{-0.8}^{+1.3}$ \\ & $-$ & & $1.09_{-0.54}^{+0.87}$ & $3.1_{-1.0}^{+1.6}$ & $3.0_{-1.5}^{+2.2}$& $3.5_{-1.0}^{+1.1}$\\ \hline
		2005-BLG-351 & $+$ & 8.5 (fixed) & $3.1_{-0.6}^{+1.1}$& $1.14_{-0.26}^{+0.25}$  & $1.4_{-0.7}^{+1.5}$& $2.0_{-0.8}^{+1.1}$\\ \hline
		BLG172.8.614 & $+$ & 8.5 (fixed) & $0.86_{-0.28}^{+0.56}$ & $1.79_{-0.61}^{+0.62}$ & $1.2_{-0.8}^{+1.4}$& $1.3_{-0.7}^{+1.1}$\\ \hline
		  {2002-BLG-061} & $+$ &   {$0.98_{-0.37}^{+1.09}$ } & $4.6_{-1.9}^{+3.7}$ & $0.80_{-0.29}^{+0.64}$ & $5.3_{-2.4}^{+3.9}$& $3.1_{-1.0}^{+1.1}$ \\ & $-$ & & $5.4_{-2.9}^{+5.5}$ & $0.71_{-0.28}^{+0.60}$ & $5.1_{-2.5}^{+3.6}$& $2.8_{-0.9}^{+1.1}$\\ \hline
		2008-BLG-096 & $+$ & $2.8_{-1.1}^{+2.4}$ &  $0.80_{-0.37}^{+1.11}$ & $1.9_{-0.8}^{+1.2}$ & $2.1_{-1.1}^{+2.3}$& $2.0_{-0.8}^{+1.2}$\\ \hline
		2005-BLG-372 & $+$ & 8.5 (fixed) & $2.34_{-0.51}^{+0.65}$ & $4.32_{-0.49}^{+0.48}$ & $1.8_{-0.9}^{+1.5}$& $4.3_{-1.2}^{+1.1}$\\ \hline
		2005-BLG-086 & $+$ & 8.5 (fixed) & $0.43_{-0.14}^{+0.24}$ & $2.76_{-0.77}^{+0.76}$ & $1.2_{-0.6}^{+1.4}$& $2.8_{-1.1}^{+1.2}$\\ \hline
		  {2005-BLG-020} & $+$ &   {8.5 (fixed) } & $0.74_{-0.19}^{+0.24}$ & $2.89_{-0.53}^{+0.51}$ & $0.8_{-0.4}^{+0.6}$& $2.5_{-0.8}^{+1.1}$\\ & $-$ & & $0.47_{-0.13}^{+0.18}$ & $2.43_{-0.53}^{+0.51}$ & $0.6_{-0.3}^{+0.5}$& $1.5_{-0.7}^{+1.1}$\\ \hline
		2006-BLG-251 & $+$ & 8.5 (fixed) &  $1.02_{-0.27}^{+0.38}$ & $2.00_{-0.046}^{+0.048}$ & $1.1_{-0.6}^{+0.7}$& $3.5_{-1.0}^{+1.0}$\\ \hline
		  {2008-BLG-013} & $+$ &   {$4.9_{-1.7}^{+2.7}$ } & $0.44_{-0.22}^{+0.44}$ & $4.3_{-1.5}^{+2.2}$ & $2.5_{-1.2}^{+1.7}$& $5.4_{-1.4}^{+1.0}$ \\ & $-$ & & $0.26_{-0.17}^{+0.41}$ & $4.4_{-1.7}^{+2.6}$ & $2.2_{-1.1}^{+1.5}$& $5.3_{-1.4}^{+1.0}$\\ \hline
	\end{tabular}
	}
		\label{tab:results table}
\end{table*}



\section{Results and Conclusion}
\label{conclusion}
In this work we compared the OGLE and GAIA catalogues to investigate possible additional information from the GAIA catalogue for the microlensing events. The extra data from the GAIA catalogue are the distance and the proper motions of the stars. Moreover, for the microlensing events with few years time-separation between the OGLE and GAIA observations, we may resolve the lens and source stars. Our simulation showed that for $0.26\%$ of microlensing events, GAIA can resolve the lens and source stars with angular separation $>400$~mas. 

Here, we investigate $13$ events with significant microlensing-parallax signature that has been studied in \cite{2016MNRAS.458.3012W}. They have estimated the mass of the lens from the parallax parameter  based on MCMC analysis of the microlensing light curve, on the other hand from blending, they have concluded that this sub-set of events have dark lenses. For this set of events, we identified the distance and proper motion of the source stars in the GAIA catalogue. For $6$ events out of $13$ microlensing events, we have found the negative parallax, reported in the 
GAIA catalogue \cite{2018A&A...616A...9L} where in future data releases of GAIA, the parallax parameter might be well measured. We fixed the distance of these source stars at $8.5$kpc in the Galactic bulge. For $7$ microlensing events we had the distance of the source stars from GAIA catalogue. Also for all the events, GAIA reported the proper motion of the source stars. 

We used the distribution function for the velocity of lenses from the Galactic model and find the likelihood values for the mass and the distance of the lenses. Since we imposed two extra constraint of the distance and the proper motion of the source stars, we could have better constraint on the mass and distance of lenses. We compared our results with that of \cite{2016MNRAS.458.3012W} in Table (\ref{tab:results table}). For those events with a fixed distance of $D_s = 8.5$ kpc, since we have the measurement of the source-proper motion, we obtained a different mass compare to the previous study. For a fixed distance of source and distance of lens from equation (\ref{eq:etime}), the lens mass relates to the relative proper motion as $M\propto \mu^2$. This rough relation is almost consistent with our numerical calculation in Table (\ref{tab:results table}).  

\begin{figure}
	\includegraphics[width=1\columnwidth]{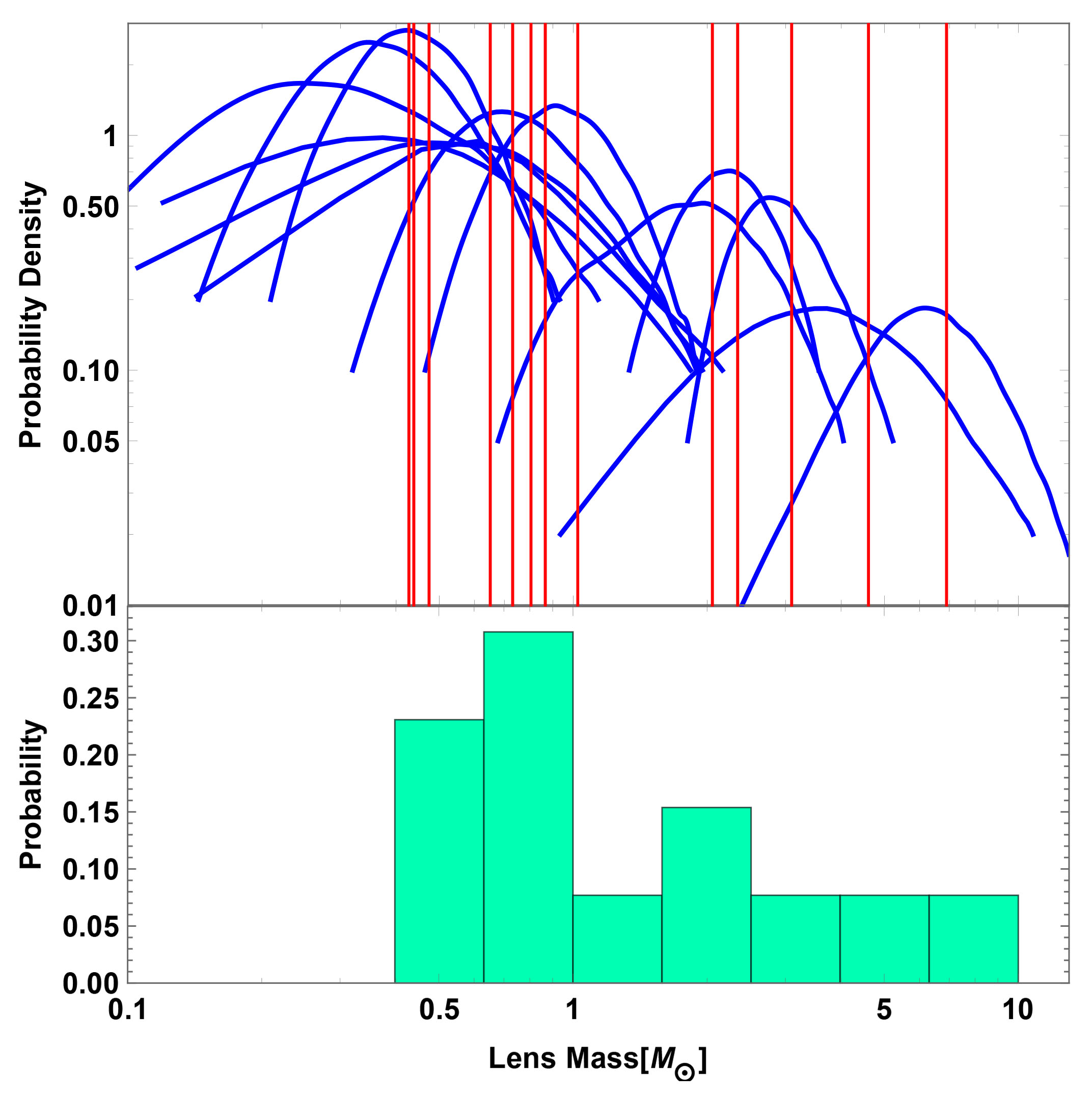}
	\caption{Histograms for probability distribution of the mass of 13 microlensing events with dark lens. Vertical red lines (in the upper panel) represents the median of each probability function representing the mass of a microlensing event. Bottom panel shows a normalised histogram of medians.}
	\label{fig:his}
\end{figure}

For $7$ microlensing events where the distance of source stars are given with the GAIA catalogue, unlike to the study in \cite{2016MNRAS.458.3012W}, the distance of the source stars are less than $8$kpc. Some of them are located in the disk or at the edge of the Galactic bulge. The masses inferred  from our analysis are smaller than that in \cite{2016MNRAS.458.3012W}. Figure (\ref{fig:his}) represents the histogram of the lens mass from this analysis for $13$ microlensing events. 
 Further studies on cross-matching between the GAIA and OGLE catalogues can reveal the mass function of single lenses from the brown dwarfs to the compact objects as the neutron stars and black holes\cite{2013arXiv1309.6635K,1998ApJ...499..367B}.  

\section*{Acknowledgements}
This research was supported by Sharif University of Technology's Office of Vice
President for Research under Grant No. G950214.
This work has made use of data from the European Space Agency (ESA) mission
{\it Gaia} (\url{https://www.cosmos.esa.int/gaia}), processed by the {\it Gaia}
Data Processing and Analysis Consortium (DPAC,
\url{https://www.cosmos.esa.int/web/gaia/dpac/consortium}). Funding for the DPAC
has been provided by national institutions, in particular the institutions
participating in the {\it Gaia} Multilateral Agreement. We would like to thank referee for his/her useful comments improving this work. 

\bibliographystyle{ws-ijmpd}
\bibliography{references}

\end{document}